\documentclass{article}

\usepackage{arxiv}

\usepackage[utf8]{inputenc} 
\usepackage[T1]{fontenc}    
\usepackage{hyperref}       
\usepackage{url}            
\usepackage{booktabs}       
\usepackage{amsfonts}       
\usepackage{nicefrac}       
\usepackage{microtype}      
\usepackage{lipsum}		
\usepackage{graphicx}
\usepackage{natbib}
\usepackage{doi}
\usepackage[ruled]{algorithm2e}
\usepackage{multicol}
\usepackage{amsmath}

\title{Toward Real-world Image Super-resolution via Hardware-based Adaptive Degradation Models}

\author{ \href{https://orcid.org/0000-0002-8836-4217}{\includegraphics[scale=0.06]{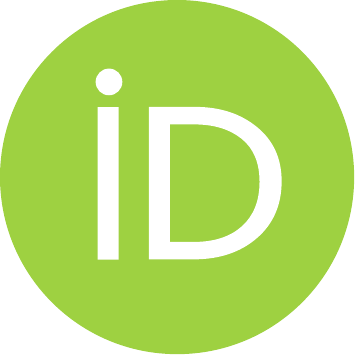}\hspace{1mm}Rui Ma} \\
	Center for Personalized Health Monitoring (CPHM), Institute for Applied Life Sciences (IALS)\\
	Department of Electrical and Computer Engineering, University of Massachusetts Amherst, MA, U.S.A.\\
	\texttt{ruima@umass.edu} \\
	\And
	{\hspace{1mm}Johnathan Czernik} \\
	Center for Personalized Health Monitoring (CPHM), Institute for Applied Life Sciences (IALS)\\
	Department of Mechanical and Industrial Engineering, University of Massachusetts Amherst, MA, U.S.A.\\
	\texttt{jczernik@umass.edu} \\	
	\And
	\href{https://orcid.org/0000-0003-3515-2840}{\includegraphics[scale=0.06]{orcid.pdf}\hspace{1mm}Xian Du}\thanks{} \\
	Center for Personalized Health Monitoring (CPHM), Institute for Applied Life Sciences (IALS)\\
	Department of Mechanical and Industrial Engineering, University of Massachusetts Amherst, MA, U.S.A.\\
	\texttt{xiandu@umass.edu} \\
}

\renewcommand{\shorttitle}{\textit{arXiv} Template}

\hypersetup{
pdftitle={A template for the arxiv style},
pdfsubject={q-bio.NC, q-bio.QM},
pdfauthor={David S.~Hippocampus, Elias D.~Striatum},
pdfkeywords={First keyword, Second keyword, More},
}

\begin{document}
\maketitle

\begin{abstract}
      Most single image super-resolution (SR) methods are developed on synthetic low-resolution (LR) and high-resolution (HR) image pairs, which are simulated by a predetermined degradation operation, e.g., bicubic downsampling. However, these methods only learn the inverse process of the predetermined operation, so they fail to super resolve the real-world LR images; the true formulation deviates from the predetermined operation. To address this problem, we propose a novel supervised method to simulate an unknown degradation process with the inclusion of the prior hardware knowledge of the imaging system. We design an adaptive blurring layer (ABL) in the supervised learning framework to estimate the target LR images. The hyperparameters of the ABL can be adjusted for different imaging hardware. The experiments on the real-world datasets validate that our degradation model can estimate LR images more accurately than the predetermined degradation operation, as well as facilitate existing SR methods to perform reconstructions on real-world LR images more accurately than the conventional approaches. 
\end{abstract}

\keywords{Real-world Super-resolution \and Image degradation \and Supervised learning }

\section{Introduction}
Image super resolution (SR) is used to reconstruct the high-resolution (HR) images from the given low-resolution (LR) images, playing an essential role in computer vision tasks. Image SR applies to numerous scenarios, including enhancing the details and photorealism of an image \cite{ledig_photo-realistic_2017}, breaking the sensor limitation of the imaging system \cite{wronski_handheld_2019}, etc. Recently, the advancement of deep learning facilitates the accuracy of SR algorithms. Many deep learning-based SR algorithms \cite{dong_image_2015}\cite{lim_enhanced_2017}\cite{kim_accurate_2016} are proposed by taking advantage of large-scale datasets. However, the performance of these state-of-the-art methods deteriorates when encountering the real-world LR inputs, even though they perform well on the synthesized, e.g. bicubic-downsampled, LR images. 

In overcoming this issue, some recent approaches use the method of collecting high-quality pairs of real-world LR and HR examples to learn their SR models. However, such an acquisition process remains to be challenging due to the spatial misalignment. Consequently, the SR models trained by real-world datasets cannot be generalized into the various scenarios of applications, e.g., medical images \cite{zhang_identify_2019}, micron or nano scale images \cite{yan_consistent_2021}, due to the lack of the diversity within the high-quality real-world datasets. Another solution to this issue is the unsupervised method to learn a degradation model first \cite{wang_unsupervised_2021}, then the following SR models are trained in the supervised manner based on the learned degradation model. One of the challenges in such methods arises from the preservation of the image content across different scales while learning the degradation model \cite{son_toward_2021}. The other challenge of these unsupervised methods is the universality to the numerous data from different imaging systems. The degradation model trained on one dataset may not be applicable in another dataset acquired by a different imaging system. Furthermore, most datasets do not provide real LR images for training the degradation model. This puts into question the universality of the degradation model, which is essential for applying the model into different real-world scenarios. 

To address these challenges, we design a supervised framework to learn a degradation model from HR to LR images. The model can be generalized into other imaging acquisition systems by adapting the hyper-parameters of the model. Similar to the previous supervised frameworks \cite{lim_enhanced_2017}, we train a degradation model to simulate real-world LR images. However, rather than updating the whole model during training, we design an Adaptive Blurring Layer (ABL) to preserve the hardware information of the imaging system that will be fixed during training. The ABL simulates the blurring process of the real-world imaging hardware, and the parameters can be adjusted based on the different hardware specifications. The contributions of this paper can be organized threefold:
\begin{itemize}
  \item We present a novel supervised learning framework to learn the unknown degradation process from HR to LR images.
  \item The degradation model can be extended into different imaging acquisition systems by adjusting the hyper-parameters of the model.
  \item The proposed method can be easily integrated with existing SR frameworks and achieve better results on real-world images.
\end{itemize}

\section{RELATED WORK}
This section surveys current solutions for the blind super-resolution problem.

To solve the reduced performance issue, researchers proposed two categories of methods. The first category of methods have high-quality HR and LR image pairs acquired by real-world imaging systems. In \cite{kohler_toward_2019}, the authors introduce the Super-Resolution Erlangen (SupER) database, the first comprehensive laboratory SR database of all-real acquisitions with pixel-wise ground truth. The data is obtained via hardware binning, and covers difficult conditions like local object motion and photometric variations. In \cite{cai_toward_2019}, the authors build a real-world super-resolution (RealSR) dataset where paired LR-HR images on the same scene are captured by adjusting the focal length of a digital camera. An image registration algorithm is developed to progressively align the image pairs at different resolutions. 
In \cite{zhang_zoom_2019}, the authors collect the paired real-world dataset, SR-RAW, via optical zoom. To deal with the mild misalignment, the authors introduce a novel contextual bilateral loss (CoBi). These aforementioned methods only solved the specific imaging system scenarios, and cannot be generalized due to the difficulties of data acquisition.

The second category of methods simulate the real-world LR images from the given HR images. Some researchers try to include all blurring kernels into the degradation model. In \cite{zhang_learning_2018}, the authors introduce a dimensionality stretching strategy that facilitates the network to handle multiple and even spatially variant degradations with respect to blur kernel and noise. By taking the concatenated LR image and degradation maps as input, the authors claim the SR model will learn to choose the appropriate degradation maps. In \cite{gu_blind_nodate}, instead of inputting all the degradation maps, the image-specific blur-kernel is iteratively learned and updated by the corrector. Then, the SR model follows the high-level architecture of SRResNet \cite{ledig_photo-realistic_2017} and extends it to handle multiple kernels by spatial feature transform (SFT) layers. However, using the predetermined downsampling operator (bicubic downsampling) as initialization constrains the simulated LR images, limiting deviations from the known formulations. In \cite{zhou_kernel_2019}, a kernel modeling super-resolution network (KMSR) is proposed where the simulated LR images are generated by applying a specific blur-kernel to HR images, which is chosen from a predetermined kernel pool. However, the SR model still needs some real-world unpaired LR and HR images to estimate the blur-kernel; it is also challenging to build the predetermined kernel pool. Some researchers try to learn the downsampling model under the Generative Adversarial Network (GAN) \cite{goodfellow_generative_2014} framework to imitate the distribution of real-world LR images. In \cite{ferrari_learn_2018}, the authors propose a two-stage process which firstly trains a high-to-low GAN to learn how to degrade and downsample the high-resolution images. During training, only unpaired high and low-resolution images are required. Once this is achieved, the output of this network is used to train a low-to-high GAN for image super-resolution, using paired low- and high-resolution images. Although the degradation model from the training HR to LR datasets is learned well, it cannot be generalized to other datasets due to the divergent imaging acquisition systems. Similarly, in \cite{maeda_unpaired_2020}, the author produces pseudo-clean LR images as the intermediate products from ground-truth HR images, which are then used to train the SR network in a paired manner. In \cite{bell-kligler_blind_2019}, the authors introduce "KernelGAN" – an image-specific Internal-GAN, which estimates the SR kernel (downscaling kernel) that best preserves the distribution of patches across scales of the LR image. However, the training process is time consuming because each image-specific SR kernel has to be trained by each LR image. On the other hand, various regularization terms need to be applied to constrain the kernel space. Other researchers use unsupervised methods to estimate the degradation process. In \cite{wang_unsupervised_2021}, the authors first learn abstract representations to distinguish the various degradations in the representation space rather than the explicit estimation in the pixel space. Then they introduce a Degradation-Aware SR (DASR) network with flexible adaption to various degradations based on the learned representations. A contrastive loss is used to conduct unsupervised degradation representation learning by contrasting positive pairs against negative pairs in the latent space. However, the defined contrastive loss is based on the assumption that any different images are incurring different degradation processes even with the same imaging acquisition system. In \cite{son_toward_2021}, the authors propose an effective way of imitating the real-world LR images of an unknown distribution. They use low-frequency loss (LFL) and adaptive data loss (ADL) to simulate accurate and realistic LR images from HR images without relying on any predetermined downsampling operators. 
\begin{figure}[h]
\vspace{.3in}
\centerline{\includegraphics[scale=0.35]{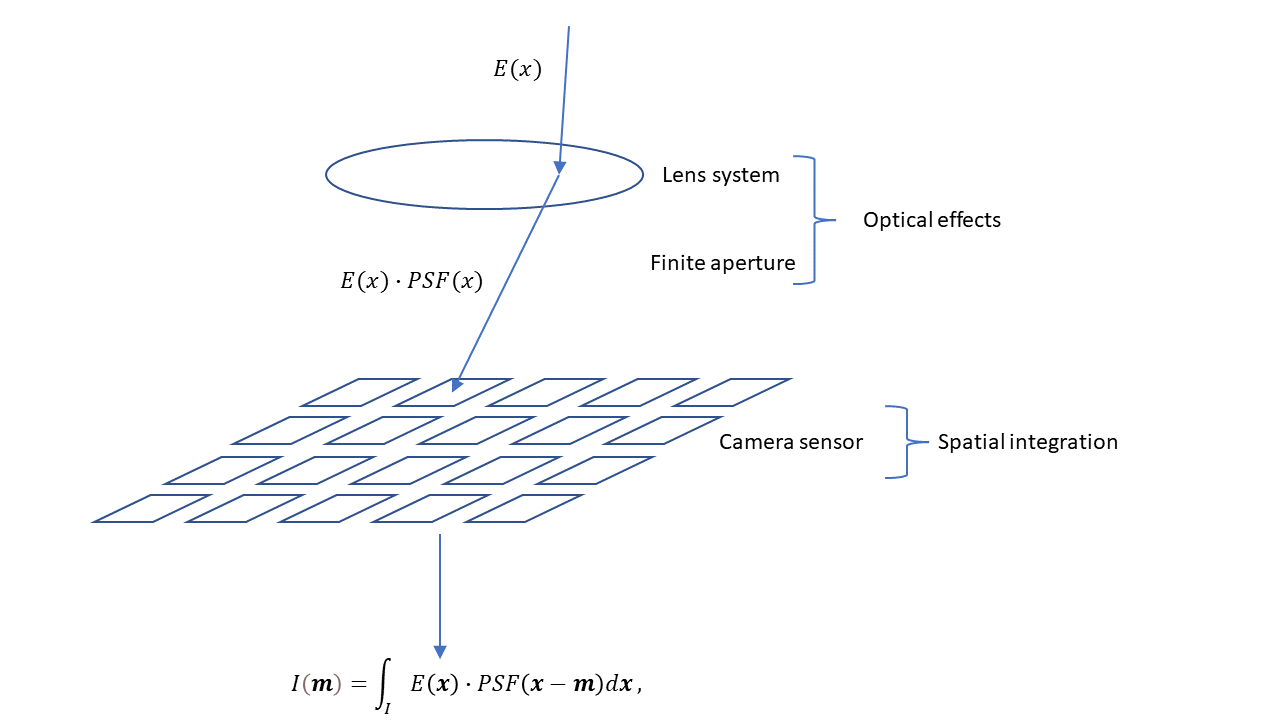}}
\vspace{.3in}
\caption{Digital image acquisition workflow}
\label{fig:1}
\end{figure}

\section{METHOD}
Section 3.1 creates the relationship between HR and LR spaces from the optical imaging perspective. Section 3.2 presents our solution for simulating the LR images.

\subsection{Problem Formulation}
\begin{figure*}[h]
\centerline{\includegraphics[width=0.8\textwidth]{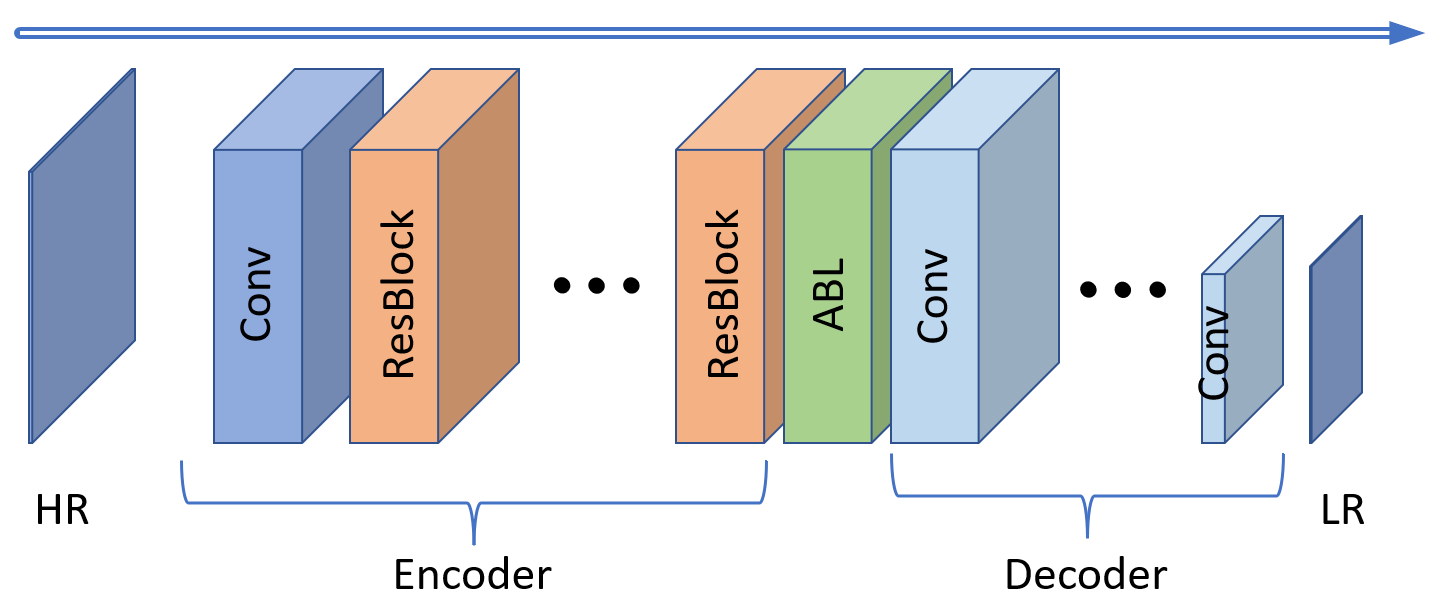}}
\caption{The architecture of the degradation learning network}
\label{fig:2}
\end{figure*}

As shown in Figure \ref{fig:1}, the digital image is acquired through the lens system and the camera sensor. When a point source goes through the lens system, there will be an area response on the focal plane of the camera. The point spread function (PSF) describes this area. The degree of the blurriness comes from the PSF of the lens system. A limited number of pixels of the camera sensor will discretize the continuous response on the focal plane - the pixel resolution comes from the camera sensor. By modeling this degradation process, we have:
\begin{equation}\label{eqn:1}
    I(\mathbf{m}) = \int_{I}{E(\mathbf{x})\cdot PSF(\mathbf{x}-\mathbf{m})d\mathbf{x}},  
\end{equation}

where $\mathbf{m}=(m,n)$ is a vector in $\mathbb{Z}^2 $ containing the pixel coordinates, $E(\cdot)$ is the continuous irradiance light-field that would have reached the image plane of I, $PSF(\cdot)$ is the point spread function of the camera, and $\mathbf{x}=(x,y) \in \mathbb{R}^2$ are the coordinates in the continuous image plane of $I$. Applying equation \ref{eqn:1} to the real-world scenario, we have two imaging systems to capture the HR image and LR image on the same field of view (FOV). Using $Lo$ denote a low-resolution image and $Hi$ to denote a high-resolution image corresponding to a finer grid, we have 
\begin{equation}\label{eqn:2}
    Lo(\mathbf{m}) = \int_{Lo}{E(\mathbf{x})\cdot PSF_{Lo}(\mathbf{x}-\mathbf{m})d\mathbf{x}}.  
\end{equation}
and \\
\begin{equation}\label{eqn:3}
    Hi(\mathbf{p}) = \int_{Hi}{E(\mathbf{z})\cdot PSF_{Hi}(\mathbf{z}-\mathbf{p})d\mathbf{z}}.  
\end{equation}
Here, $\mathbf{m}=(m,n)$ and $\mathbf{p}=(p,q)$ are vectors in $\mathbb{Z}^2 $ containing the pixel coordinates of the LR and HR images, $\mathbf{x}=(x,y) \in \mathbb{R}^2$ are coordinates in the continuous image plane of $I$. Because LR and HR images have the same FOV, the image planes $E(\cdot)$ should be the same. Furthermore, in the case of optical zoom, $PSF_{Hi}$ is a narrower version of $PSF_{Lo}$. Suppose the same FOV is taken by the same camera, the optical zoom-in factor of the HR image from LR image is $\alpha$, we have:
\begin{equation}\label{eqn:4}
    PSF_{Hi}(x)=\alpha PSF_{Lo}(\alpha x).
\end{equation}
Substituting \ref{eqn:2}, \ref{eqn:3} and \ref{eqn:4}, the HR and LR images are mainly transformed by blurring and subsampling:
\begin{equation}\label{eqn:5}
    Lo[\mathbf{m}] = \sum_{p} Hi[\mathbf{p}]k[\alpha \mathbf{m}-\mathbf{p}],
\end{equation}
where $k$ is a discrete blurring kernel. According to \cite{michaeli_nonparametric_2013}, the optimal kernel in real-world scenario is usually narrower than the PSF. Therefore, modeling the blurring kernel gives rise to the blind super-resolution problem, namely, to super resolve the LR image under the unknown blurring kernel from HR to LR.

\subsection{Learning to Downsample}

Based on the formulation in section 3.1, in order to simulate the accurate LR images, the variance of the imaging acquisition system should be considered, because different sensors and lens will affect the degradation process. Most unsupervised methods are trying to mimic this process, while the supervised methods are learning the exact models from the ground-truth HR and LR pairs. Therefore, the supervised methods will be more accurate. However, there is a significant challenge in the supervised methods as the learned degradation model cannot be extended into other datasets which are acquired from different imaging devices. Therefore, we proposed a supervised method to learn the degradation model, and the model can be generalized into other datasets by adapting the hyper-parameters of the model. 

As shown in Figure 2, our designed architecture of the degradation network takes a gray scale HR image as the input. Next, going through the stacks of residual blocks, the input image will be represented by the high-dimensional feature maps. Then, the ABL will blur the feature maps channel-wise. Last, the blurred high-dimensional feature maps will go through the decoder and form the LR image. We will explain this framework in further details below.

\subsubsection{High-dimensional representation}
As shown in section 3.1, the image degradation process is not convolved by a simple discretization of the PSF function, and cannot be described by a simple convolution kernel. Therefore, to represent the non-linear degradation process, we first design some stacks of residual blocks to encode the input image into the high-dimensional feature map. Different from the original ResNet \cite{he_deep_2016}, we get removed the batch normalization layers inside the residual blocks (See Figure 3) as \cite{lim_enhanced_2017} and \cite{nah_deep_2017} did. The reason is that batch normalization layers normalize the features, and then eliminate the range flexibility of networks. Furthermore, simpler architecture can reduce the computational complexity and the memory cost. As shown in Figure 2, the encoder module converts the input image into the high-dimensional representation.
\begin{figure}[h]
\vspace{.3in}
\centerline{\includegraphics[scale=0.3]{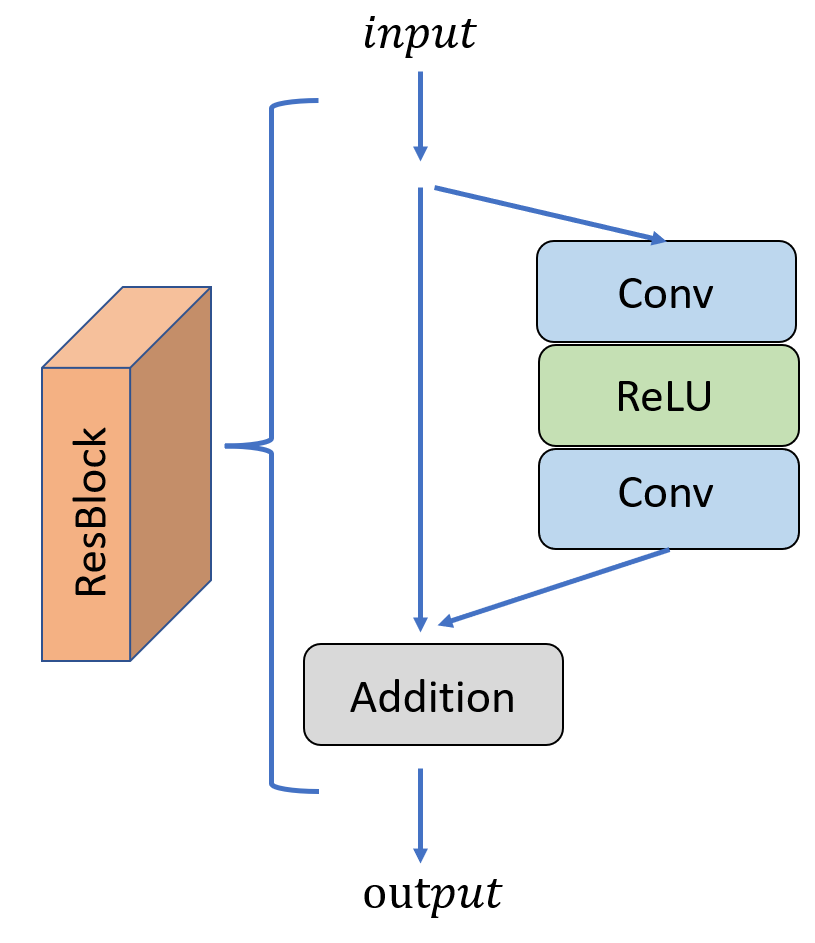}}
\vspace{.3in}
\caption{The architecture of the residual block (ResBlock)}
\label{fig:3}
\end{figure}

\subsubsection{Adaptive blurring layer}

\begin{figure}[h]
\vspace{.3in}
\centerline{\includegraphics[scale=0.3]{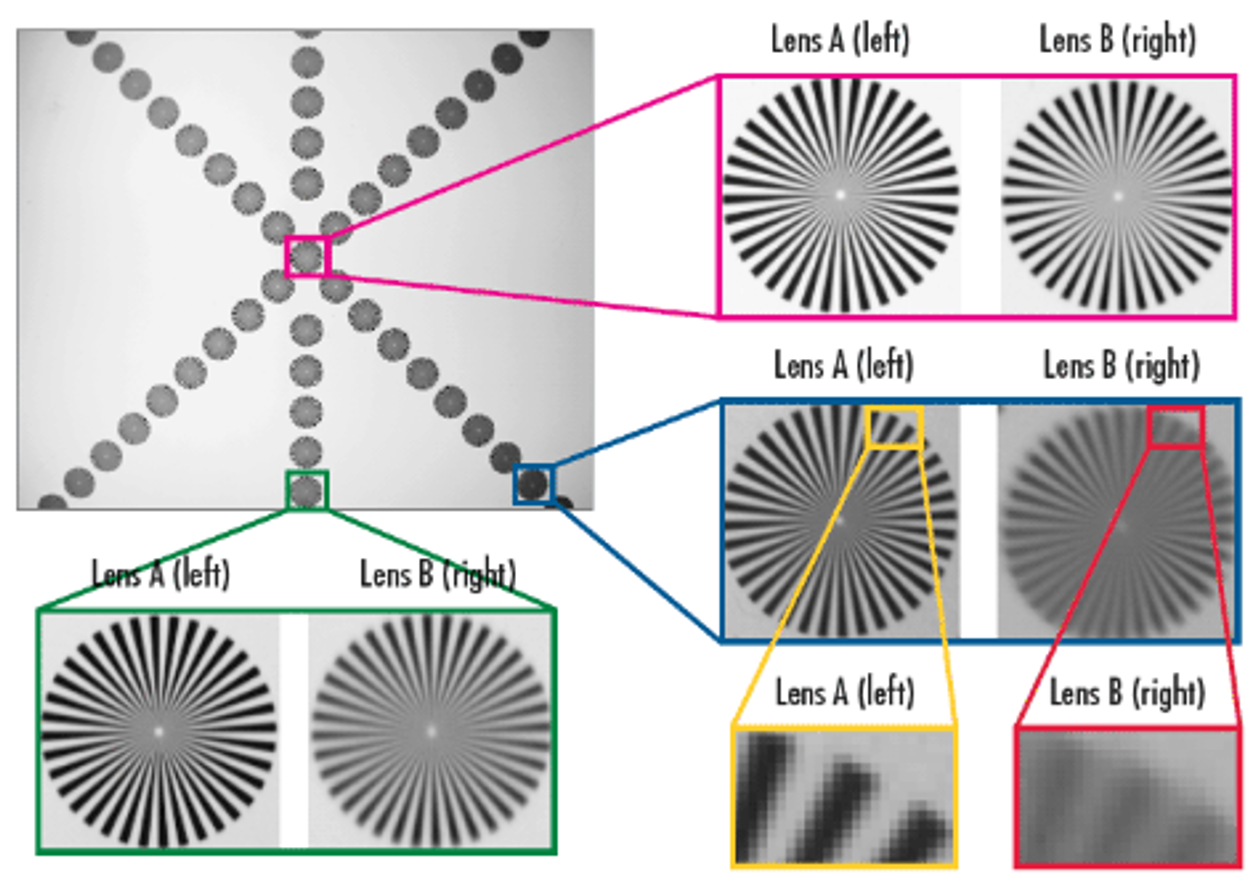}}
\vspace{.3in}
\caption{A star target is imaged with two lenses (A and B) with the same focal length, f/\#, field of view, and camera sensor. The superiority of lens A is apparent in all areas but is not pronounced along the edge and in the corner of the image \cite{noauthor_lens_nodate}}
\label{fig:4}
\end{figure}
Given the high-dimensional representation of the input HR image, we propose a channel-wise blurring layer to model the degradation process from HR to LR images. Intuitively, the high-dimensional representation extracts multiple features from the original HR image. To mimic the degradation process in real-world scenarios, two aspects should be considered. First, each feature can incur a slightly different degradation process even though the whole image is acquired by the same imaging hardware. The reason is that in  real-world imaging systems, the camera sensor and the lens system are both unideal. For example, as shown in Figure 4, both of the lenses show a better contrast level in the center than the edges and corners. Furthermore, the high-dimensional representation may extract more latent features that incur different degradation processes. Also, the degradation process from the same imaging hardware should follow a basic rule to discriminate the different hardware, e.g., the two lenses A and B in Figure 4. 

Considering the aforementioned hardware complexity, we design a Gaussian Mixture Model (GMM) as the blurring kernel for each channel of the high-dimensional representation. As shown in Figure 5, each channel of the high-dimensional representation convolves with multiple weighted anisotropic Gaussian kernels (four in Figure 5), and then adds up the convolution results into a single channel again. The anisotropic Gaussian kernels have different "directions" and "scales" corresponding to the uncertainty of real-world imaging systems. However, the change of the scales will be controlled within a small range, corresponding to the consistency of the current imaging hardware. Next, to embed the GMM into the supervised learning framework, we design a weight for every anisotropic Gaussian kernel. The weights for each channel of the high-dimensional representation add up to one for intensity consistency. Consequently, the weights of each anisotropic Gaussian kernel will be decided by the learning process. 

\begin{figure*}[h]
\centerline{\includegraphics[width=0.8\textwidth]{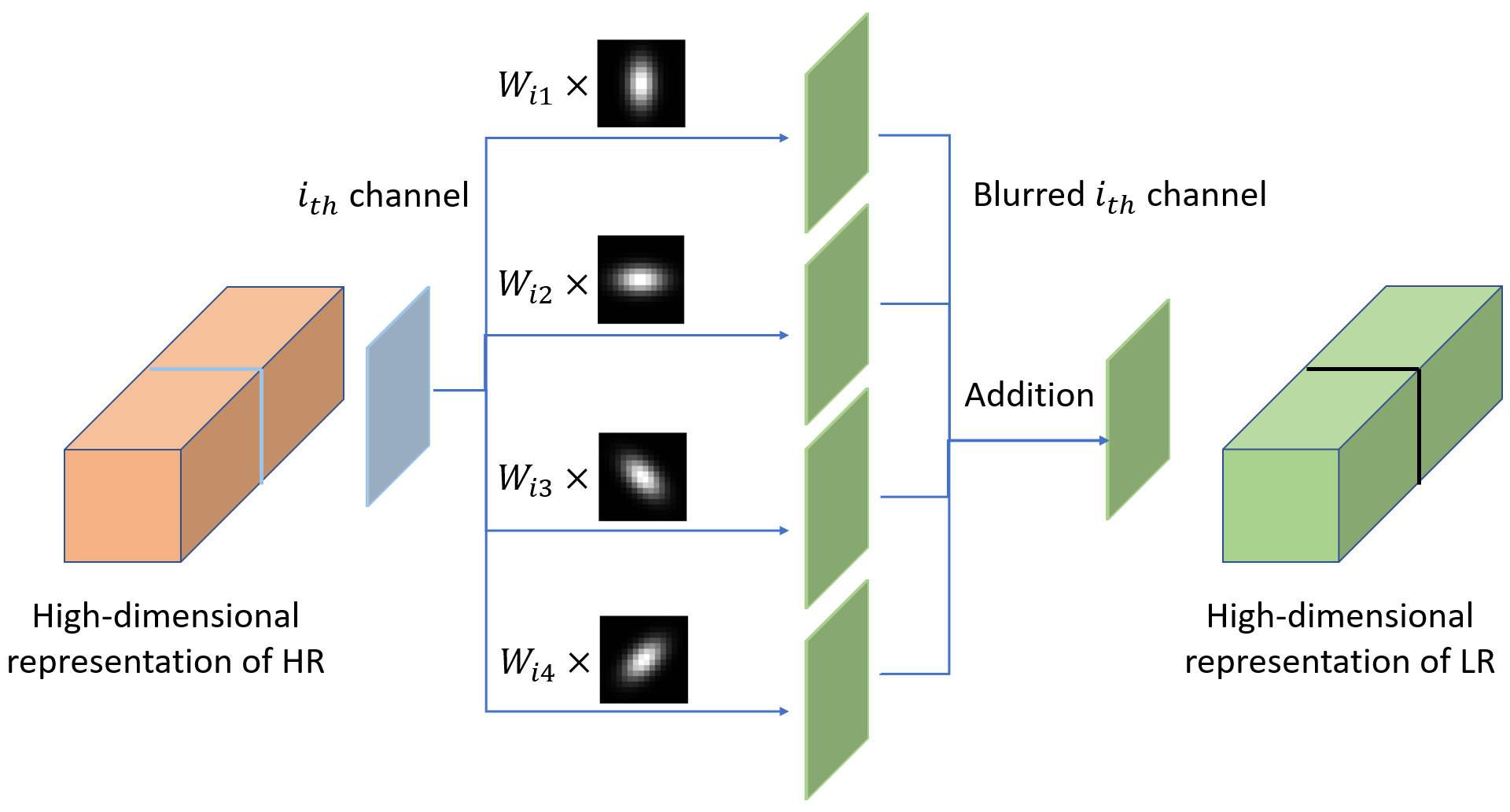}}
\caption{The architecture of the ABL}
\label{fig:5}
\end{figure*}

Algorithm 1 shows the method to get the fixed size Gaussian kernel. First, we choose the fixed region of interest (ROI) for all the kernels, e.g., $\{(x,y)|-4<x<4,-4<y<4\}$ in Algorithm 1. Second, we project the fixed size of pixels onto this region with a scale factor. There will be a sub-region for each pixel, and the pixel value is the integration of the PDF along with the sub-region. 

The reasons in choosing GMM are as follows: 
\begin{enumerate}
  \item Anisotropic Gaussian function is a good way to simulate PSF \cite{baker_limits_2002}.
  \item Each channel of the feature map can be blurred differently by the weighted GMMs, which simulates the real-world imaging process.
  \item The "scales" of the GMMs can be changed when encountering a different imaging system, e.g., from lens A to lens B in Figure 4.
\end{enumerate}

\begin{algorithm}
\DontPrintSemicolon

\KwIn{ROI of the Gaussian function, default=$\{(x,y)|-4<x<4,-4<y<4\}$; kernel size, default=$16 \times 16$.}
\KwOut{The fixed size Gaussian $kernel$.}
Initialization: $x=-4,scale\_factor=16/(4-(-4)),m = 1, n = 1, kernel = zeros(16,16)$\

\While{$x \leq 4$}
{
  set $y=-4$\;
  \While{$y \leq 4$}
  {
    kernel(m,n) = CDF of Gaussian in $\{(i,j)|x<i<x+1/scale\_factor,y<j<y+1/scale\_factor\}$\;
    n = n + 1\;
    y = y + 1/scale\_factor\;
  }
  n = 1\;
  m = m + 1\;
  x = x + 1/scale\_factor\;
}
\caption{Discrete Gaussian filter calculation}

\end{algorithm}

\subsubsection{Decoder}
After going through the ABL, the blurred high-dimensional representation of the HR image, which we call the high-dimensional representation of the LR image, will be decoded into the single channel gray scale image with the required size. We implement multiple convolution layers with the ReLU activation function to achieve the decoding process. The number of layers and stride values can be adjusted based on the downsampling factor. The advantage of this architecture is to change the image size smoothly and keep enough receptive field of each pixel of the output LR image.

\section{EXPERIMENTS}
We implement our method based on the PyTorch environment. We will release the source code and the trained models. The following are the implementation details of our method. We evaluate our proposed method on real-world paired HR and LR image datasets, SupER \cite{kohler_toward_2019}, and RealSR\cite{cai_toward_2019}. 

\subsection{Experimental Setups}
\subsubsection{Dataset}
SupER dataset is a newly proposed high-quality paired HR and LR image dataset for real-world image super-resolution tasks. The dataset is acquired by Basler acA2000-50gm COMS camera with a $f/1.8$, 16 mm fixed-focus lens in the laboratory. The dataset consists of more than $80,000$ images of 14 scenes combining different facets. We select only the "inlier" images as our training data. \\
RealSR dataset is acquired by Canon 5D3 and Nikon D810 camera with 24-105mm, f\/4.0 lens. It consists of both indoor and outdoor scenes. 
\subsubsection{Training Details}
We use the gray scale image patches of size $256\times256$ from the HR images with the corresponding LR patches of size $64\times64$ as our input. We augment the training data with random horizontal and vertical flips. All the images are rescaled into $[0,1]$ before training. We train our model with the ADAM optimizer by setting $\beta_1=0.9,\beta_2=0.999,$. The training batch size is set to 8, and the learning rate is set to $10^{-4}$. 
We select 112 HR images as training data and randomly extract 50 patches in one HR image. 50 epochs are trained for each model on NVIDIA RTX 3090 GPU. L1 loss was chosen as the loss function due to its pixel-wise nature. In addition, L2 loss does not necessarily yield better performance when considering SSIM and PSNR according to \cite{zhao_loss_2016}.

For the parameters of ABL, we set the high-dimensional representation to be 128 channels. Then, we design two groups of experiments: the first group contains four Gaussian filters with various orientations; the second group contains eight Gaussian filters with different orientations and scales. In the first group, we set the covariance matrices to be: 

\begin{equation}\label{eqn:6}
    factor \cdot \left ( R(angle)\times\begin{bmatrix} 1 &0 \\  0 &0.3  \end{bmatrix}\times R(angle)^T \right ), 
\end{equation}
where $R(\cdot)$ represents the rotation matrix and the angles are set to $[0, 45^{\circ}, -45^{\circ}, 90^{\circ}]$. The factor is fixed in each experiment of group one, and five models are trained in group one with the factor value set to $[0.25; 0.5; 1; 2; 3]$. In the second group, we set the factor to $1.2\times factor$ for the other four Gaussian filters while keep the four Gaussian filters in group one the same, i.e., the factor values of the five models in group two are set to $[0.25, 0.3; 0.5, 0.6; 1, 1.2; 2, 2.4; 3, 3.6]$.

\subsection{Evaluation on the Generated LR Images}

\subsubsection{Evaluation on SupER dataset}

The degradation model was trained on a small part of the SupER dataset (112 HR and LR image pairs). After training, we tested the models on 129 unused images. As shown in Table 1, our degradation models outperforms the bicubic downsampling method. 

Furthermore, the best result in group one is the model with the lowest factor (0.25) while the results are similar for different factors in group two. Such a result can be attributed to the fact that adding the other set of Gaussian filters with different scales gives more flexibility to the model. On the other hand, if we can select an appropriate factor value by the prior knowledge of the imaging hardware (0.25 and 0.5 in Table 1), one set of fixed scale Gaussian filters can be trained more efficiently and then give the same accurate result. Section 4.2.2 will further validate the appropriate factor value of the model using the RealSR dataset.

\begin{table}[h]
\footnotesize
\caption{Performance on SupER dataset} \label{table:1}
\begin{center}
\begin{tabular}{cc|cc}
\multicolumn{2}{c}{Group one} &\multicolumn{2}{c}{Group two} \\
\begin{tabular}{@{}c@{}}\textbf{$factor$ in} \\ \textbf{equation 6}\end{tabular} &L1 Loss&
\begin{tabular}{@{}c@{}}\textbf{$factor$ in} \\ \textbf{equation 6}\end{tabular}& L1 Loss \\
\hline\hline
0.25&    0.0088&  0.25, 0.3& 0.0107\\
0.5 &    0.0111&  0.5, 0.6 & 0.0096\\
1   &    0.0140&  1, 1.2 &   0.0097\\
2   &    0.0379&  2, 2.4 &   0.0089\\
3   &    0.0906&  3, 3.6&    0.0105\\
\hline
\multicolumn{2}{c|}{bicubic downsampling} &\multicolumn{2}{c}{0.0197} \\

\end{tabular}
\end{center}
\end{table}

\subsubsection{Evaluation on RealSR Dataset}
Table 2 and Table 3 show the performance of transferring the model into a different dataset. Given the same two groups of trained models, we test each model with different factors on the RealSR dataset. The test data taken by the Canon camera is used to evaluate the performance. After choosing the appropriate factor, the performance of the generated LR images are 30\% better. 

Furthermore, both Table 2 and 3 show that the trained model with a greater factor performs better in the RealSR dataset while the trained models with a smaller factor gives better results in both SupER and RealSR datasets. The imaging system in SupER is Basler acA2000-50gm COMS camera with a $f/1.8$, 16 mm fixed-focus lens. The imaging system in RealSR is Canon 5D3 and Nikon D810 which is not as accurate as the Basler industrial imaging systems. Therefore, the PSF of the Basler system should be narrower than the Canon one. The results are consistent with the imaging hardware of both datasets. 

\begin{table}[h]
\footnotesize
\caption{L1 Losses of Group one's models on RealSR dataset} \label{table:2}
\begin{center}
\begin{tabular}{c|ccccc}
\multicolumn{1}{c}{model} &\multicolumn{5}{c}{adjusted factor values for the current model} \\
factors& 0.25& 0.5& 1& 2& 3 \\
\hline\hline
0.25&    0.0315&    0.0292&  0.0265&  0.0252&  0.0258\\
0.5 &    0.0345&    0.0305&  0.0260&  \textbf{0.0233}&  0.0237\\
1   &    0.0468&    0.0405&  0.0324&  0.0266&  0.0259\\
2   &    0.0891&    0.0666&  0.0475&  0.0310&  0.0261\\
3   &    0.1833&    0.1275&  0.0785&  0.0451&  0.0322\\
\hline
\multicolumn{3}{c}{bicubic downsampling} &\multicolumn{3}{c}{0.0317} \\
\end{tabular}
\end{center}
\end{table}

\begin{table}[h]
\footnotesize
\caption{L1 Losses of Group two's models on RealSR dataset} \label{table:3}
\begin{center}
\begin{tabular}{c|ccccc}
\multicolumn{1}{c|}{model} &\multicolumn{5}{c}{adjusted factor values for the current model} \\
factors& 0.25& 0.5& 1& 2& 3 \\
\hline\hline
0.25&    0.0308&    0.0284&  0.0253&  \textbf{0.0232}&  0.0234\\
0.5 &    0.0349&    0.0314&  0.0278&  0.0254&  0.0255\\
1   &    0.0445&    0.0383&  0.0311&  0.0256&  0.0245\\
2   &    0.0768&    0.0615&  0.0470&  0.0332&  0.0284\\
3   &    0.1342&    0.1055&  0.0730&  0.0436&  0.0311\\
\hline
\multicolumn{3}{c}{bicubic downsampling} &\multicolumn{3}{c}{0.0317} \\
\end{tabular}
\end{center}
\end{table}
\subsection{SR performance on the real-world images}
\begin{figure*}[h]
\centerline{\includegraphics[width=\textwidth]{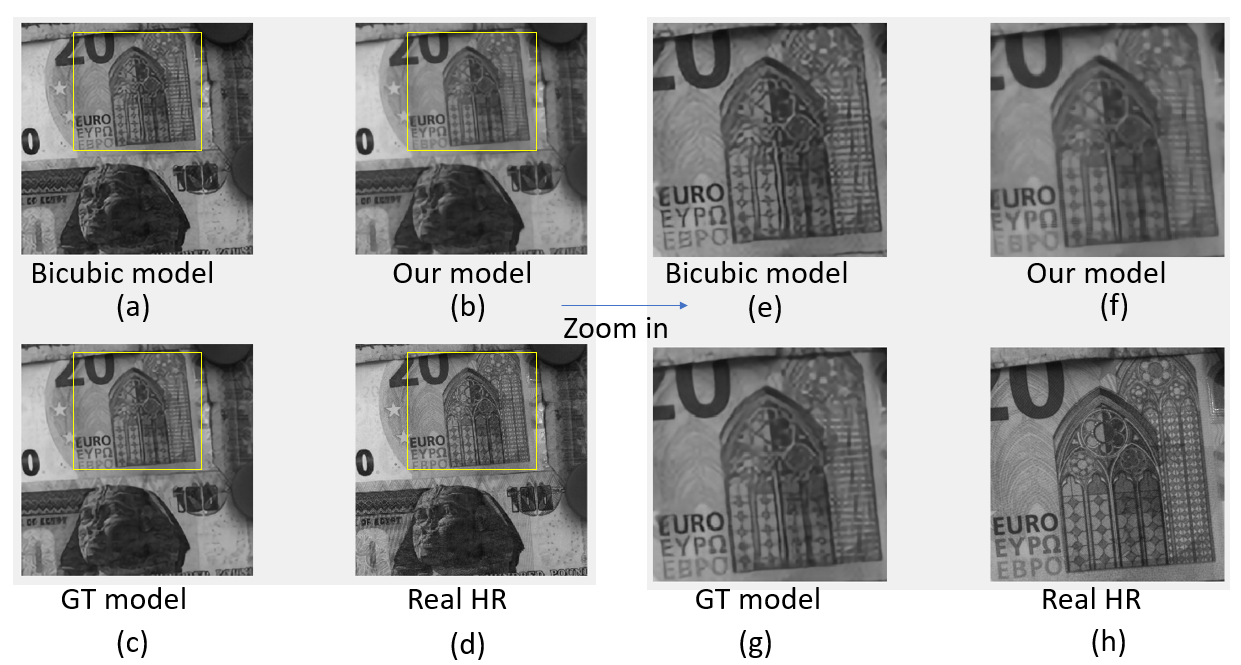}}
\caption{The visualization of SR results.}
\label{fig:6}
\end{figure*}

Given the degradation model trained from the SupER dataset, LR images can be generated for different datasets without further retraining if we know the hardware of the imaging system. We test the SR results using the SupER dataset. EDSR \cite{lim_enhanced_2017} framework is used in the experiment. First, the dataset is divided into two parts: training and testing. Second, given the training HR images, we generate two sets of LR images: bicubic downsampled LR images and LR images generated by our degradation model. Third, three SR models are trained by the two generated LR, and the real-world ground-truth LR. Last, we evaluate the performance by inputting the real-world LR images into the three models. As shown in Table 4, the SR model trained by our LR images achieves higher PSNR and SSIM than the bicubic model, and is close to the ground-truth model, which proves our degradation model improves the real-world SR performance. 

Figure 6 demonstrates the visual results comparing three SR models to ground truth HR. Figure 6.(a)(e) is a demonstration of the SR model trained with bicubically downsampled images. Figure 6.(b)(f) is a demonstration of the SR model trained with one of our group two degradation models (factor=$[2,2.4]$). Figure 6.(c)(g) represents a SR model trained with the ground-truth real-world LR and HR image pairs. Figure 6.(d)(h) is the ground-truth testing HR image. It is important to note the difference in performance between the bicubic model and our model. As shown in Figure 6.(e), the bicubic model produces the wavy lines, due to the predefined kernel of the bicubic degradation. Meanwhile, our model produces a straight line, similar to that in the ground-truth HR image (See Figure 6.(f) and Figure 6.(h)).

\begin{table}[h]
\footnotesize
\caption{Performance in real-world SR framework} \label{table:3}
\begin{center}
\begin{tabular}{c|ccc}

model trained by& PSNR& SSIM&\\
\hline\hline
bicubic downsampled LR and HR&    29.84&    0.859&  \\
LR generated by our model and HR&    31.70&  0.864\\
ground-truth LR and HR&              32.65&  0.877\\

\hline
\end{tabular}
\end{center}
\end{table}

\section{CONCLUSION}
We propose a novel supervised framework to learn the unknown degradation process from real-world HR to LR images. The learned degradation model can facilitate existing SR methods to perform more accurate reconstructions on real-world LR images than the conventional approaches. Furthermore, we design an adaptive blurring layer to preserve the information of imaging hardware, so that the learned degradation model can be extended into other imaging acquisition systems by only adapting the hyper-parameters of the model. Experiments show that our model produces more accurate LR images for SR training than the conventional bicubic downsampling on multiple real-world datasets. Existing SR method, e.g., EDSR\cite{lim_enhanced_2017}, trained on LR and HR pairs generated by our model shows a better result than the original bicubic downsampled LR and HR pairs. More experiments are underway for testing on the RealSR dataset.

\bibliography{draft} 
\bibliographystyle{unsrtnat}

\end{document}